# Searching for a standard Drake equation

José Antonio Molina Molina
DXC Technology
joseantonio2molina@gmail.com

**Abstract**

In the 20th century the scientific search for extraterrestrial intelligence began, and the Drake equation was proposed to estimate the number of extraterrestrial species humanity could attempt to detect, *N*. This paper discusses the need to update and standardize this equation. A new and more accurate expression is deduced which contains the classic equation as a particular case, and its advantages are discussed. A necessary condition is also stated for its use in scientific contexts: if *N* is defined as the total number of civilizations like ours then $N = 1$, given that we exist, and consequently the working hypothesis of the SETI project can be expressed as $N>1$. In this case, the Drake equation is being applied in a scientific way, because it is trying to test a hypothesis based on evidence.

## 1. Introduction: the classic Drake equation

In 1961 Frank Drake proposed the following equation, to try to estimate the number of civilizations, *N*, capable of making radio communications in our galaxy:

$$N = R_* f_p n_e f_l f_i f_c L \qquad (1.1)$$

where: $N$ = number of civilizations in the galaxy that emit radio signals; $R_*$ = rate of star formation; $f_p$ = fraction of them with planetary systems; $n_e$ = number of habitable planets in each planetary system; $f_l$ = fraction of them that have developed life; $f_i$ = fraction of them that have developed intelligent life; $f_c$ = fraction of civilizations that perform interstellar communications; and $L$ = length of time that these civilizations release detectable signals into space.

The Drake equation does not express any scientific discovery, but is a heuristic resource [1] and was part of the agenda of the first conference about the SETI project [2]. The high uncertainty in some of the factors implies that it cannot obtain a valid estimation for *N*, but the equation has become an important tool for SETI, because it stimulates the reflection about the conditions that make possible the existence of technological civilizations in the galaxy. The equation does not banish our ignorance, but it helps us to organize it and to delimit search strategies.

## 2. Why a standard Drake equation is necessary

A standard and consensual form of the Drake equation would be desirable to avoid the formal, interpretative and operative dispersion that has accompanied the equation throughout its almost 60 years of life. These dispersions make it difficult to compare the conclusions of some studies with those of others, and cause a huge disparity of values for $N$ in the literature. This is an evidence of subjective and non-scientific uses. If the equation is going to be used in a scientific context these problems should be reduced.

Any investigation should distinguish between the raw data, and the biases, preferences or filters applied to them. The Drake equation does not fulfill this principle. The $R_*$ factor is the rate of star formation (quantity) compatible with intelligent life (selection criterion). Therefore, some authors prefer to write $R_* f_g$, where the first factor is the quantity and the second is the filter (Shklovskii, Sagan, etc., cited by Ćirković [3]). The $n_e$ factor also mixes one datum (the number of planets for each star) with a selection criterion (planetary habitability). Even the $L$ factor is sometimes accompanied by an additional fraction, $\xi$, to denote the concept of 'communication window' [3].

The result of this is that different authors use different groups of factors, which causes that there is no a universal or a canonical Drake equation [3]. If a strategy is adopted of separating the gross quantities from the selection criteria, our uncertainties will be more precisely delineated and the possibility that all authors work with the same common form of the equation will increase -without feeling the need to alter it.

The search for extraterrestrial civilizations is a tracking operation through different levels of physical reality: stars, planets, biospheres, intelligences and civilizations. It would be desirable for the equation for finding the 'Drake number' ($N$ can be called that in this paper) to reflect these levels explicitly in their factors, differentiating them from each other. Also, it would be advisable that the equation to be expressed in a universal notation, without being redefined according to the preferences of the authors. The notation should be versatile, so it could be adapted to the purposes of each study.

Drake's equation was proposed almost 60 years ago, and the search for extraterrestrial technological activity was concentrated on radio waves [4]. Nowadays, the activity of the SETI project has been expanded: techno-signatures are searched for in the optical range [5], X-rays [6], gamma rays (Makovetskii, 1980, and Lemarchand 1994, cited by Tarter [7]), infrared light [8], or neutrinos [9]. Extraterrestrial artifacts are also sought [10], including novel proposals, such as Clarke's exo-belts in exoplanets in transit [11], and other more classic ones, such as the Dyson spheres [12] or its overcrowding, the Fermi bubbles [13].

The new era of SETI suggests expanding the interpretation of Drake's number: it could be the number of extraterrestrial civilizations with a technology capable of provoking artificial alterations that would be detectable from astronomical distances using our current technology. Possibly they will be civilizations that have initiated a space age. In short, for the purposes of this work they might be called 'astrocivilizations'.

The above decisions can reduce the formal and interpretative dispersion of the Drake equation. To reduce operational dispersion, a consensus should be established about the conditions of use of the equation in scientific contexts, also establishing the starting evidence, the hypothesis made, and the activity necessary to validate the hypothesis. This will avoid the usual subjective and non-scientific uses of the equation.

## 3. Number of astrocivilizations in the galaxy: progressive counting through levels of organization

A civilization is a complex system created by intelligence, which leaves detectable traces in its environment due to its activity. Intelligence, as it is known, only manifests itself in some living beings, and these only appear in some biospheres. The inorganic matter that serves as substrate to a biosphere is in the planets, and the planets only exist in planetary systems, which are developed around the stars.

The above reasoning suggests a structure for reality in the form of levels of organization. Our search involves counting the elements of each level and filtering the appropriate ones, multiplying them by a fraction lower than one. It can be written:

$$N = N_s F_s N_p F_p N_b F_b N_i F_i N_c F_c \qquad (3.1)$$

where: $N$ = Drake number, or number of astrocivilizations detectable at astronomical distances in the galaxy; $N_s$ = total number of stars in the galaxy; $F_s$ = stellar filter; $N_p$ = average number of planets in each planetary system; $F_p$ = planetary filter; $N_b$ = average number of biospheres on each planet; $F_b$ = biological filter; $N_i$ = average number of intelligences in each biosphere; $F_i$ = filter of intelligences; $N_c$ = average number of civilizations founded by each intelligence; and $F_c$ = filter of civilizations.

This equation is not yet the standard Drake equation that is being sought, but it will be the basis for its deduction. The factor $N_b$ may seem strange, but its inclusion does not obey only to reasons of formal regularity, and will be justified in section 5.

This equation is a particular case of a more general one. For example, if the target is to look for astrocivilizations throughout the universe, the first level to consider could be galaxies, and Equation 3.1 would include two factors at the beginning referring to galaxies, $N_g F_g$. It is also possible that other authors wish to consider levels different from those that appear in Equation 3.1, or include sublevels among them. The general equation can be expressed as follows:

$$N = \prod_{j=1}^{n} N_j F_j \tag{3.2}$$

where $N_j$ is the number of elements of level $j$, and $F_j$ is the filter applied to that level to select the elements compatible with the final objective, which is to find certain elements in level $n$.

## 4. Number of astrocivilizations as a function of their duration and the rate of star formation: obtaining a standard Drake equation

A standard Drake equation is being sought which should have the classic Equation 1.1 as a particular case. This implies that the Drake number, given by Equation 3.1, must be a function of the rate of star formation and the average duration of the astrocivilizations. The reasons why Drake used these parameters in his equation can be found, for example, in Vakoch & Dowd [2].

Table 1 represents a simplified and regular model of the galaxy, over several time units. In each unit a new generation of stars is born, some of them develop planets (letter *p*), and some of them develop biospheres (letter *b*). For simplicity, the other levels (intelligences and civilizations) are not considered. For example, *6s* symbolizes, in a given unit of time, a star of the $6^{th}$ generation; passing the time, if it develops two planets, it is denoted by *6spp*, and if it develops a biosphere, it is denoted by *6sppb*. At the end of its life, by losing its biospheres and its planets, the star will be denoted again by *6s*. In this model every star last for 8 units of time. For avoiding border effects due to limitations of the Table, it can be assumed that the observer is in the unit of time 9.

First, let's suppose the target is to get the number of stars with planets. Equation 3.1 is reduced to the following:

$$N = N_s F_s \tag{4.1}$$

In this case, the star filter, $F_s$, is simply the fraction of stars with planets. The quantity $N$ can be put as a function of the rate of star formation in the galaxy, given that:

$$N_s = R_s Age_{s\text{-}older} \tag{4.2}$$

where $R_s$ is the rate of star formation (stars per unit of time), and $Age_{s\text{-}older}$ is the age of the oldest stars in the galaxy, an amount that coincides with the number of existing star generations, by the regularity of the model. In the example of Table 1, in unit of time 9 it can be seen that $N_s = R_s Age_{s\text{-}older} = 6 \times 8 = 48$ stars. Substituting Equation 4.2 in 4.1 it results:

$$N = R_s Age_{s\text{-}older} F_s \tag{4.3}$$

Let's call $f_s$ the fraction of stars of the same generation (arising in the same unit of time) that develop planets. The target now is to frame Equation 4.3 in terms of this parameter, which is equivalent to putting $F_s$ as a function of $f_s$. As $f_s$ refers to the stars of a generation, let's relate $F_s$ to the generations of stars. Let's call $N_{s-p}$ the number of stars that have planets, then, $F_s = N_{s-p}/N_s$.

If a star is too young, it will not yet have a planetary system; if it is too old, she will have lost it (if it ever had it). Suppose that the minimum age is $age_{s-min}$ and the maximum is $age_{s-max}$; therefore, the age of the star, $age_s$, must meet $age_{s-min} < age_s \leq age_{s-max}$. In the set of units of time, or of generations, given by $Age_{s-older}$, the number of generations that fulfill the previous condition is the quantity $age_{s-max} - age_{s-min} \equiv L_s$, where it has been called $L_s$ the length of time in the one that a star can have a planetary system. In the model of Table 1 it can be seen that $age_{s-min} = 2$ and $age_{s-max} = 7$, therefore $L_s = 5$; the generations of stars that meet that their age is greater than 2 and less than or equal to 7 are given by the set of 5 elements {3, 4, 5, 6, 7}.

In any unit of time, therefore, there are $L_s$ generations of stars that *can* have planets, because they have an appropriate age to have them. In $L_s$ generations of stars there are $L_s R_s$ stars. In each generation, how many stars have planets? That information is provided by the fraction $f_s$. So:

$$F_s = N_{s-p}/N_s = L_s R_s f_s / R_s Age_{s-older} \tag{4.4}$$

This results in the sought relationship between $F_s$ and $f_s$:

$$F_s = (L_s / Age_{s-older}) f_s \tag{4.5}$$

When replacing in Equation 4.3, it results:

$$N = R_s f_s L_s \tag{4.6}$$

This equation can be applied to the galactic model of Table 1 and it will be obtained the expected result: $N = R_s f_s L_s = 6 \times (5/6) \times 5 = 25$ stars with planets.

Now the target is to get the number of planets that have developed biospheres. From Equations 3.1 and 4.6 it results:

$$N = R_s f_s L_s N_p F_p \tag{4.7}$$

$N_p$ is the average number of planets in each planetary system, observed in any 'photo' of the galaxy. Let $n_p$ be the average number of planets in each planetary system calculated on the stars of a same generation. Because of the regularity of the model it is indifferent if $N_p$ is calculated for a set of stars of different ages, or for stars of the same age, therefore, $N_p \equiv n_p$.

In this case, the $F_p$ filter matches the fraction of planets that have developed biospheres in a planetary system, that is, $F_p = n_{p-b}/n_p$, where $n_{p-b}$ is the average number of planets with biospheres, and $n_p$ the average number of planets in a planetary system. The numerator and the denominator can be multiplied by $N_{s-p}$, the number of stars with planets, and it results:

$$F_p = n_{p\text{-}b}/n_p = n_{p\text{-}b}N_{s\text{-}p}/n_pN_{s\text{-}p} \tag{4.8}$$

Note that the numerator is the multiplication of the number of stars with planets ($N_{s\text{-}p}$), by the number of planets with biospheres for each star ($n_{p\text{-}b}$). Therefore, it is the total number of planets with biospheres in the galaxy. For its part, the denominator is the total number of planets in the galaxy. It is possible to write the denominator like this: $n_pN_{s\text{-}p} = n_pL_sR_sf_s$

For a planetary system to develop a biosphere it must have a certain age range. This is equivalent to saying that the star must have an age within a certain range. Suppose that the difference between the maximum age and the minimum age of the star compatible with a biosphere is the length of time $L_p$. This amount matches the number of generations of stars that can have biospheres in a given unit of time.

The total number of planets with biospheres in the galaxy ($n_{p\text{-}b}N_{s\text{-}p}$) is equal to the number of stellar generations that have an appropriate age to be able to have biospheres ($L_p$), times the number of stars that are born in each generation ($R_s$), times the fraction of them that have planets ($f_s$), times the number of planets in each star ($n_p$), times the fraction of them that can develop biospheres, which can be called $f_p$. That is to say:

$$F_p = L_pR_sf_sn_pf_p/n_pL_sR_sf_s = (L_p/L_s)f_p \tag{4.9}$$

If this equation is incorporated into Equation 4.7, it gives:

$$N = R_sf_sL_sN_pF_p = R_sf_sL_sn_p(L_p/L_s)f_p = R_sf_sn_pf_pL_p \tag{4.10}$$

This expression can be applied to the model in Table 1. It is already known that $R_s= 6$ stars per unit of time and $f_s = 5/6$. To calculate $n_p$ it should be remembered its definition: it is the *average* number of planets in a planetary system, therefore $n_p = (3+2+2+1+1)/5 = 9/5$. Calculating $f_p$ is determined by the average number of biospheres in a planetary system, which is $(2+1+0+0+0)/5 = 3/5$. Dividing this by the average number of planets in each planetary system, $n_p$, results in $f_p = 1/3$. On the other hand, $L_p = 2$ units of time. When we substitute these values in Equation 4.10 the result is that $N = R_sf_sn_pf_pL_p = 6\times(5/6)\times(9/5)\times(1/3)\times 2 = 6$ planets with biospheres. This result matches what was expected (see Table 1).

The relation between $F_p$ and $f_p$ (Equation 4.9), is analogous to that between $F_s$ and $f_s$ (Equation 4.5). If a similar analysis is performed for the rest of the levels, it will result in the characteristic time length of one level cancelling the next, and in the end there would be only the length of time $L_c$, corresponding to the civilizations. On the other hand, it is true that $L_s>L_p>L_b>L_i>L_c$, so the last one is the one that imposes the greatest restriction on Drake's number.

After performing the same analysis across all levels, Equation 3.1 becomes the relationship that was being sought:

$$N = R_s f_s n_p f_p n_b f_b n_i f_i n_c f_c L \qquad (4.11)$$

where the subscript $c$ of the length of time $L$ has been deleted, since there is no danger of ambiguity because the rest of the time lengths have been canceled.

The above equation gives the number of Drake, $N$, as a function of the rate of star formation in the galaxy, $R_s$, and the average duration of astrocivilizations, $L$, and can be considered a standardized form of the Drake equation.

Equation 4.11 is a particular case of a more general one, when instead of the proposed levels (stars, planets, biospheres, intelligences and civilizations) an arbitrary set of $n$ levels is proposed. It can be proved from the Equation 3.2.

$$N = \prod_{j=1}^{n} N_j F_j = R_1 L_0 F_1 \prod_{j=2}^{n} N_j F_j = R_1 L_0 \left(\frac{L_1}{L_0}\right) f_1 \prod_{j=2}^{n} n_j \left(\frac{L_j}{L_{j-1}}\right) f_j$$

$$N = R_1 f_1 L_1 \prod_{j=2}^{n} n_j f_j \prod_{j=2}^{n} \left(\frac{L_j}{L_{j-1}}\right) = R_1 f_1 L_1 \prod_{j=2}^{n} n_j f_j \frac{L_n}{L_1}$$

In the first equality the number of elements of the first level has been put in function of the emergency rate of them, $R_1$, giving $N_1 = R_1 L_0$ equivalently to Equation 4.2. The factor $L_0$ is the age of the oldest elements of level 1. Each $N_j$ coincides with $n_j$, which is the average number of elements of level $j$, as justified after Equation 4.7. Each filter $F_j = (L_j / L_{j-1}) f_j$, by analogy with Equations 4.5 and 4.9, where each $f_j$ is the filter applied to the elements of level $j$ and each $L_j$ is the necessary duration that the elements of level $j$ must have for the level $j+1$ to exist. Since each level is contained within the previous one, and only exists within it, the causal connectivity implies that $L_{j-1} > L_j > L_{j+1}$.

The above development gives rise to what it can be called the standard Drake equation in its most generic form, provided that the conditions of regularity implicit in all this development are assumed:

$$N = R_1 f_1 L_n \prod_{j=2}^{n} n_j f_j \qquad (4.12)$$

To avoid writing the sequence product symbol, and remembering Einstein's notation for summations over repeated indexes, it may be suggested to use a notation that replaces it, such as the cross ($\times$), or midpoint ($\cdot$). The above equation would look like this:

$$N = R_1 f_1 (\cdot n_j f_j) L_n \qquad (4.13)$$

An even more synthetic way of writing it is by defining a parameter, $R$, which would be the emergence rate of astrocivilizations in the galaxy, $R = R_s f_s n_p f_p n_b f_b n_i f_i n_c f_c$, or $R = R_1 f_1 (\cdot n_j f_j)$, so that the equation for the number of Drake is $N = RL$.

Table 2 contains a summary of the standard Drake equation found in the previous development.

A necessary condition to be met by the standard Drake equation is that it must contain the classical equation. Equation 1.1 can be obtained from Equation 4.12 by doing: $R_1 \equiv R_*$, $f_1 \equiv f_p$, $L_n \equiv L$, $n_2 \equiv n_e$, $f_2 = 1$, $n_3 = 1$, $f_3 \equiv f_l$, $n_4 = 1$, $f_4 \equiv f_i$, $n_5 = 1$, $f_5 \equiv f_c$. A more accurate equivalence can be found by breaking down the filters in each of the selection criteria of the next section. The equivalences are shown in Table 3.

## 5. Suggestions for the selection of filters in the standard Drake equation, and justification of the $n_b$ factor

The versatility of the standard Drake equation is related to the filters applied in each level. Each filter is the product of a series of fractions, and each fraction is a selection criterion. Each fraction is identified with a subscript that reveals which filter it belongs to, and which selection criterion it obeys. On the other hand, the last criterion or fraction, in each filter, is a bridge with the next level, and determines the value of the number of elements in that next level: $f_{s-p}$ implies a definition of a 'planet', which will determine the value of the number $n_p$, etc.

Some suggestions for the selection criteria for the filters of the standard Drake equation could be as follows:

(i) $f_s = f_{s-gh} f_{s-h} f_{s-p}$ where $f_{s-gh}$ = galactic habitability; $f_{s-h}$ = stellar habitability and $f_{s-p}$ = stars with planets;

(ii) $f_p = f_{p-r} f_{p-h} f_{p-b}$ where $f_{p-r}$ = rocky planets; $f_{p-h}$ = planetary habitability and $f_{p-b}$ = planets with biospheres or 'biosystems';

(iii) $f_b = f_{b-cb} f_{b-s} f_{b-i}$ where $f_{b-cb}$ = carbon-based biospheres; $f_{b-s}$ = surface biospheres and $f_{b-i}$ = biospheres with intelligent species;

(iv) $f_i = f_{i-a} f_{i-nr} f_{i-c}$ where $f_{i-a}$ = anthropic-type intelligences; $f_{i-nr}$ = intelligences that reach a Neolithic or agricultural revolution and $f_{i-c}$ = intelligences that create civilizations (urban revolution);

(v) $f_c = f_{c-e} f_{c-t} f_{c-d}$ where $f_{c-e}$ = expansive civilizations; $f_{c-t}$ = technological civilizations, or those that reach an industrial revolution and $f_{c-d}$ = civilizations detectable at astronomical distances.

This is not the place to discuss each of the factors, filters and fractions; nevertheless, it is necessary to justify $n_b$, the number of biospheres on each planet. It is usually thought that a planet has only one biosphere ($n_b = 1$), but this does not always have to be the case.

For example, the Earth itself could harbor 'hidden' or 'rare' biospheres, composed of microorganisms [14]. On the other hand, planets with captured rotation (Proxima b) or with inland oceans (Europe, Titan...) could develop greatly differentiated biospheres with separate evolutions. Also, if each value of the $n_p$ factor corresponds to a planet with its satellites, then even for a gas giant it could happen that $n_b > 1$, if it has several moons that have developed biospheres. In any case, it would be more correct to use the word 'biosystem' - understood as an organization of living beings and inert matter characterized by a particular and independent trophic chain. On Earth, marine life and terrestrial life would be differentiated surface biosystems.

## 6. Use of the standard Drake equation: the evidence $N = 1$

The Drake equation has been used as a tool to reflect on the conditions that make extraterrestrial civilizations possible in the galaxy, which helps us to design search strategies. It cannot give us an exact value for $N$, but it dissects this number transforming it into several factors, which helps us to identify and to limit our uncertainties, challenging us to dissipate them.

However, throughout its history the equation has also been used to quantify the number of civilizations in the galaxy. To avoid the uncertainties in the factors, every author makes their own choices based on their own analysis. Even if these analyses are well reasoned, they are often affected by the particular views of the author related to the existence of extraterrestrial civilizations, and these determine the final values chosen for the factors. This fact explains the huge and absurd disparity of results for $N$ in the literature. The values cover up to 8 orders of magnitude – an unprecedented dispersion in other scientific equations [15].

It is necessary to establish a distinction between what a scientific use of the equation means, and other uses too contaminated by the imagination and by the idiosyncrasies or partial points of view of each author. That limit is related to the difference between objective and subjective point of view. The objective point of view here is what it is known with certainty – the one that all authors accept without discussion. The question that arises from that is: what it is known with certainty, in this matter?

In this galaxy it is possible that there exists an intelligent species like ours, capable of founding a technological civilization like ours, which generates techno-signatures detectable from astronomical distances like ours. If we continue to grow, our impact on our natural environment will increase and we will be even more detectable. By examining the trends of our technological development, we can imagine the production of new anthropogenic techno-signatures. We do not know anything else.

In the previous paragraph, a piece of evidence is established that can be used to define a 'standard Drake number': it is the number of technological civilizations in the galaxy similar to ours ('anthropic-type'), that provoke techno-signatures similar to those that we provoke now, or in the immediate future, and in such a way that they are detectable with our current technology. It is known for sure that such civilizations *can* exist in the galaxy, because it is known that *there is* one – ours. Therefore, the Drake number, when defined in this way, is not a conjecture: it is a means of quantifying the scientific evidence, and its value, for the time being, is $N = 1$.

From there, our scientific hypothesis consists of moving from the singular ($N = 1$) to the plural ($N > 1$): as we exist, our existence is possible; if it has been possible here, perhaps it has also been possible in other regions of the galaxy, because there is no reason to think that we occupy a privileged region. This is part of the scientific activity, which tries to identify regularities in nature [16]: there is no reason to exclude the phenomena of life and intelligence in that exploration based on the search of regularities as if they were a kind of universal asymmetry. It is not even necessary to cite Copernicanism or the mediocrity principle, which are sometimes misused [17].

If Drake's number and equation are interpreted in this way, then they are valid tools for science, because we are starting from an objective basis and shared by the entire scientific community (this basis is the 'Drake evidence', $N = 1$), and our target is scientific because it is based on a valid hypothesis (the 'Drake hypothesis', $N > 1$). This opens before us several lines of research related to each of the factors of the equation. We should determine the set of conditions that have made the existence of our technological civilization possible, then estimate how frequently those conditions occur in the galaxy, and in what regions they might be found. Also, the techno-signatures that we produce should be identified and their remote detectability evaluated, because other anthropic-type civilizations could be producing them as well. All scientific projects that follow this line of work, which can be called 'SETI standard', will be comparable, and that could help to quantify our progress.

The search for astrocivilizations similar to ours does not derive from a 'terrestrial chauvinism' [18], but is a necessary condition to guarantee the scientific nature of the search. This does not mean that other lines of research should be impeded: the existence of very advanced civilizations can be imagined, such as those of Type II or III on the Kardashev scale [19], and we can try to search for them. However, it should be remembered that, for these civilizations, $N = 0$, and therefore the search no longer starts from the standpoint of evidence.

If the SETI project should be part of a worldwide effort, as requested by Carl Sagan in 1982 and supported by several Nobel Prize winners [20], we must ask ourselves what line of work will have the most support from the international scientific community – the one that it is based on evidence, or any of those that are based on speculation?

# 7. Conclusions

This work has suggested finding a standard form for the Drake equation and has provided a concrete example. It has also stated a minimum condition to guarantee its use in scientific contexts. Table 2 shows a summary of this work.

The formulation of the proposed equation is consistent with the heuristic development undertaken: an exploration along several levels of organization (stars, planets, biospheres, intelligences and civilizations). At each level, two separate operations are performed: the elements are counted and filtered to select those that are compatible with the search. These operations are expressed in Equation 4.11 or its generalization, Equation 4.12. In the classical Drake equation, the tracking operation along organizational levels is not as evident, and gross quantities are mixed with selection criteria.

Likewise, Equation 4.11 is easily scalable or applicable to other sets of levels; Equation 4.12 is its most general formulation, valid for any set of levels. Some authors may need to consider intermediate levels to those proposed here, or may wish to start with another first level: instead of stars, galaxy zones, galaxies, groups or clusters of galaxies, etc. could be proposed.

The versatility of the equation is also related to the filters. There is a filter for each level (stellar filter, planetary filter...), each filter is a product of fractions, and each fraction represents a specific selection criterion (see Section 5). Criteria could be added or deleted without changes in the general formulation of Equation 4.11, which would remain the same. The classic Drake equation presents a less versatile structure.

This work has also defended the need to use the standard Drake equation as part of a line of work of the SETI project characterized by the objectivity, rigor and consensus of the scientific community. That consensus can only be reached if we start from an evidence, and *we are that evidence*. For civilizations like ours, or anthropic-type ones, and only for these, it is true that $N = 1$. From this evidence can be formulated the hypothesis that there are other civilizations similar to ours, and therefore, $N > 1$. This would be a necessary requirement to use the standard Drake equation in conditions that guarantee scientific rigor, because we are looking for something that *we know may exist*.

**Acknowledgments**

The author thanks the anonymous referees.

# Tables

## Table 1: Galactic model for the deduction of the standard Drake equation.

| 6 | 7 | 8 | 9 | 10 | 11 | 12 |
|---|---|---|---|---|---|---|
|  |  |  |  | 10s | 10s | 10spp |
|  |  |  |  | 10s | 10s | 10spp |
|  |  |  |  | 10s | 10s | 10sppp |
|  |  |  | 9s | 9s | 9s | 9s |
|  |  |  | 9s | 9s | 9sp | 9sp |
|  |  |  | 9s | 9s | 9sp | 9sp |
|  |  |  | 9s | 9s | 9spp | 9spp |
|  |  |  | 9s | 9s | 9spp | 9spp |
|  |  |  | 9s | 9s | 9sppp | 9sppp |
|  |  | 8s | 8s | 8s | 8s | 8s |
|  |  | 8s | 8s | 8sp | 8sp | 8sp |
|  |  | 8s | 8s | 8sp | 8sp | 8sp |
|  |  | 8s | 8s | 8spp | 8spp | 8spp |
|  |  | 8s | 8s | 8spp | 8spp | 8sppb |
|  |  | 8s | 8s | 8sppp | 8sppp | 8spppbb |
|  | 7s | 7s | 7s | 7s | 7s | 7s |
|  | 7s | 7s | 7sp | 7sp | 7sp | 7sp |
|  | 7s | 7s | 7sp | 7sp | 7sp | 7sp |
|  | 7s | 7s | 7spp | 7spp | 7spp | 7spp |
|  | 7s | 7s | 7spp | 7spp | 7sppb | 7sppb |
|  | 7s | 7s | 7sppp | 7sppp | 7spppbb | 7spppbb |
| 6s | 6s | 6s | 6s | 6s | 6s | 6s |
| 6s | 6s | 6sp | 6sp | 6sp | 6sp | 6sp |
| 6s | 6s | 6sp | 6sp | 6sp | 6sp | 6sp |
| 6s | 6s | 6spp | 6spp | 6spp | 6spp | 6spp |
| 6s | 6s | 6spp | 6spp | 6sppb | 6sppb | 6spp |
| 6s | 6s | 6sppp | 6sppp | 6spppbb | 6spppbb | 6sppp |
| 5s | 5s | 5s | 5s | 5s | 5s | 5s |
| 5s | 5sp | 5sp | 5sp | 5sp | 5sp | 5s |
| 5s | 5sp | 5sp | 5sp | 5sp | 5sp | 5s |
| 5s | 5spp | 5spp | 5spp | 5spp | 5spp | 5s |
| 5s | 5spp | 5spp | 5sppb | 5sppb | 5spp | 5s |
| 5s | 5sppp | 5sppp | 5spppbb | 5spppbb | 5sppp | 5s |
| 4s | 4s | 4s | 4s | 4s | 4s |  |
| 4sp | 4sp | 4sp | 4sp | 4sp | 4s |  |
| 4sp | 4sp | 4sp | 4sp | 4sp | 4s |  |
| 4spp | 4spp | 4spp | 4spp | 4spp | 4s |  |
| 4spp | 4spp | 4sppb | 4sppb | 4spp | 4s |  |
| 4sppp | 4sppp | 4spppbb | 4spppbb | 4sppp | 4s |  |
| 3s | 3s | 3s | 3s | 3s |  |  |
| 3sp | 3sp | 3sp | 3sp | 3s |  |  |
| 3sp | 3sp | 3sp | 3sp | 3s |  |  |
| 3spp | 3spp | 3spp | 3spp | 3s |  |  |
| 3spp | 3sppb | 3sppb | 3spp | 3s |  |  |
| 3sppp | 3spppbb | 3spppbb | 3sppp | 3s |  |  |
| 2s | 2s | 2s | 2s |  |  |  |
| 2sp | 2sp | 2sp | 2s |  |  |  |
| 2sp | 2sp | 2sp | 2s |  |  |  |
| 2spp | 2spp | 2spp | 2s |  |  |  |
| 2sppb | 2sppb | 2spp | 2s |  |  |  |
| 2spppbb | 2spppbb | 2sppp | 2s |  |  |  |
| 1s | 1s | 1s |  |  |  |  |
| 1sp | 1sp | 1s |  |  |  |  |
| 1sp | 1sp | 1s |  |  |  |  |

**Table 2: Summary of the standard Drake equation, and condition for its use in scientific contexts.**

| SETI standard project | | |
|---|---|---|
| **Evidence** | **Hypothesis** | **Testing the hypothesis** |
| $N = 1$, 'we are here': there is an anthropic-type civilization | $N > 1$, 'we are not alone': there are other anthropic-type civilizations | Search for anthropic-type techno-signatures, standard Drake equation |
| **Standard Drake equation** | | |
| Generic form: | $N = R_1 f_1 L_n \prod_{j=2}^{n} n_j f_j$ | Notation suggestion: $N = R_1 f_1 (\cdot n_j f_j) L_n$ <br> Abbreviated form: $N = RL_n$, con $R = R_1 f_1 (\cdot n_j f_j)$ |
| Common form (our galaxy)*: | $N = R_s f_s n_p f_p n_b f_b n_i f_i n_c f_c L$ <br> Abbreviated form: $N = RL$, con $R = R_s f_s n_p f_p n_b f_b n_i f_i n_c f_c$ <br> * Levels = {stars (s), planets (p), biospheres (b), intelligences (i), civilizations (c)} | |
| **Description of the factors** | | |
| $N$ = Standard Drake number, or number of anthropic-type astrocivilizations that cause techno-signatures which are detectable with our current technology <br> $L$ = Length of time that anthropic-type techno-signatures remain detectable <br> $R$ = Emergency rate of anthropic-type astrocivilizations in the galaxy | | |
| $R_s$ = Rate of star formation <br> $n_p$ = Number of planets for each star <br> $n_b$ = Number of biospheres for each planet <br> $n_i$ = Number of intelligent species for each biosphere <br> $n_c$ = Number of civilizations for each intelligent species | | $f_s$ = Stellar filter <br> $f_p$ = Planetary filter <br> $f_b$ = Biological filter <br> $f_i$ = Filter of intelligences <br> $f_c$ = Filter of civilizations |

**Table 3: Equivalences between the standard Drake equation and the classical Drake equation**

| Standard Drake equation | Equivalence with the classic Drake equation | Standard Drake equation | Equivalence with the classic Drake equation |
|---|---|---|---|
| $R_s f_{s-g} f_{s-h}$ | $R_*$ | $f_{b-i}$ | $f_i$ |
| $f_{s-p}$ | $f_p$ | $n_i f_{i-a} f_{i-n} f_{i-c}$ | 1 |
| $n_p f_{p-r} f_{p-h}$ | $n_e$ | $n_c$ | 1 |
| $f_{p-b}$ | $f_l$ | $f_{c-e} f_{c-t} f_{c-d}$ | $f_c$ |
| $n_b f_{b-cb} f_{b-s}$ | 1 | $L$ | $L$ |